\documentclass[aps,prx,twocolumn,amsfonts,amssymb,amsmath,nofootinbib,superscriptaddress]{revtex4-2}

\usepackage[utf8]{inputenc}
\usepackage[T1]{fontenc}
\usepackage[english]{babel}

\usepackage{graphicx}
\usepackage{xspace}
\usepackage[separate-uncertainty]{siunitx}
\usepackage{mathtools}
\usepackage{hyperref}
\usepackage[capitalize]{cleveref}

\binoppenalty=\maxdimen
\relpenalty=\maxdimen

\DeclareSIUnit\year{yr}
\DeclareSIUnit\day{d}
\DeclareSIUnit\atomicmassunit{u}

\DeclareMathOperator{\hc}{h.c.}
\DeclareMathOperator{\tr}{tr}

\newcommand*{\DM}{\ensuremath{\text{DM}}\xspace}
\newcommand*{\LQCD}{\ensuremath{\Lambda_\text{QCD}}\xspace}
\newcommand*{\Ncopy}{\ensuremath{\mathcal{N}}\xspace}
\newcommand*{\ZN}{\ensuremath{\mathbb{Z}_\Ncopy}\xspace}
\newcommand*{\etal}{\emph{et al.}\xspace}

\begin{document}
%

%
%
\title{Prospects of nuclear-coupled-dark-matter detection via correlation spectroscopy of \texorpdfstring{I$_2^+$}{I2+} and \texorpdfstring{Ca$^+$}{Ca+}}
\author{Eric Madge}
\email{eric.madge-pimentel@weizmann.ac.il}
\affiliation{Department of Particle Physics and Astrophysics, Weizmann Institute of Science, Rehovot 761001, Israel}
\author{Gilad Perez}
\email{gilad.perez@weizmann.ac.il}
\affiliation{Department of Particle Physics and Astrophysics, Weizmann Institute of Science, Rehovot 761001, Israel}
\author{Ziv Meir}
\email{ziv.meir@weizmann.ac.il}
\affiliation{Department of Physics of Complex Systems, Weizmann Institute of Science, Rehovot 761001, Israel}
\begin{abstract}
    The nature of dark matter~(DM) and its interaction with the Standard Model~(SM) is one of the biggest open questions in physics nowadays. The vast majority of theoretically motivated Ultralight-DM (ULDM) models predict that ULDM couples dominantly to the SM strong/nuclear sector.
    This coupling leads to oscillations of nuclear parameters that are detectable by comparing clocks with different sensitivities to these nature's constants. 
    Vibrational transitions of molecular clocks are more sensitive to a change in the nuclear parameters than the electronic transitions of atomic clocks. Here, we propose the iodine molecular ion, I$_2^+$, as a sensitive detector for such a class of ULDM models. The iodine's dense spectrum allows us to match its transition frequency to that of an optical atomic clock~(Ca$^+$) and perform correlation spectroscopy between the two clock species. With this technique, we project a few-orders-of-magnitude improvement over the most sensitive clock comparisons performed to date. We also briefly consider the robustness of the corresponding ``Earth-bound'' under modifications of the \ZN-QCD axion model.
\end{abstract}
\maketitle
%

%
\section{Introduction}
%

Roughly \SI{80}{\percent} of the matter content of our Universe consists of dark matter~(DM).
Despite ample evidence for its existence from astrophysical and cosmological observations, little is known about its nature.
One well-motivated candidate for DM is a bosonic field with mass,~$m_\DM$, in the range of \SI{e-22}{\eV} to a few \si{\eV},%
\footnote{We use natural units with $\hbar=c=1$ throughout most of this article.}
the so-called ultralight DM~(ULDM).
It is characterized by a large occupation number, so that ULDM can be described as a coherently oscillating background field, oscillating at its Compton frequency, $2 \pi f_C = m_\DM$, with an amplitude of $\sqrt{2 \rho_\DM}/m_\DM$, where $\rho_\DM = \SI{0.4}{\GeV\per\cubic\centi\meter}$ is the local DM density~\cite{McMillan:2011wd}.

Well-motivated models of ULDM include the axion~\cite{Preskill:1982cy,Abbott:1982af,Dine:1982ah,Hook:2018dlk,DiLuzio:2020wdo} of quantum chromodynamics~(QCD), the dilaton~\cite{Arvanitaki:2014faa} (though see Ref.~\cite{Hubisz:2024hyz}), the relaxion~\cite{Banerjee:2018xmn,Chatrchyan:2022dpy}, and possibly other forms of Higgs-portal models~\cite{Piazza:2010ye}. 
All of these predict that the ULDM would couple dominantly to the Standard Model (SM) QCD sector, the quarks, and the gluons, leading to oscillations of nuclear parameters~\cite{Flambaum:2006ip,Damour:2010rm,Damour:2010rp} such as the mass of the proton. 

Scalar ULDM generically couples linearly to the hadron masses, whereas pseudoscalar ULDM such as the QCD-axion couples quadratically to them, see, e.g., Ref.~\cite{Kim:2022ype} (see also Refs.~\cite{Kim:2023pvt,Beadle:2023flm} for induced subleading couplings to electrodynamics). 
Furthermore, one can construct a broad class of natural ULDM models where the leading DM interaction with the SM fields is quadratic~\cite{Banerjee:2022sqg}.

Variations of SM parameters can be searched for by comparing the rates of two clocks (e.g., quantum transitions or resonators) that exhibit different dependence on the parameters in question~\cite{Damour:2010rp,Arvanitaki:2014faa,Stadnik:2015kia}.
Laboratory limits on these variations have been obtained from various clock-comparison experiments based on atomic or molecular spectroscopy as well as cavities and mechanical oscillators (see Refs.~\cite{Uzan:2010pm,Safronova:2017xyt,Antypas:2020rtg,Antypas:2022asj} for a review).

Atomic-clock comparisons based on electronic transitions can probe DM coupling to the strong/nuclear sector, to leading order, only indirectly via their dependence on the reduced mass and the charge radius (see Ref.~\cite{Banerjee:2023bjc} for a recent discussion).
Measurements of atomic hyperfine transitions, as well as comparison to optical or acoustic resonators, allow probing DM interactions with the strong/nuclear sector through oscillations of the nuclear parameters.
However, hyperfine-transition frequencies are much smaller than electronic-optical frequencies, making their sensitivity much smaller. A similar argument for reduced sensitivity also holds to the dependence of energy levels to the nuclear parameters through the reduced mass or even the charge-radius~\cite{Banerjee:2023bjc}.

Direct sensitivity to the hadronic parameters can be obtained from molecules, as they further feature rotational and vibrational degrees of freedom with transitions that directly depend on the proton mass.
This makes molecules a sensitive detector to DM that couples to the SM’s strong sector.
Several molecular clock experiments have been conducted to place constraints on linear drifts of nuclear parameters, where the currently strongest molecular-clock-based bound of $\partial_t \log(\mu) \lesssim \SI{e-14}{\per\year}$ is obtained from comparing a transition between nearly degenerate vibrational levels in different electronic potentials of KRb to a Cs clock~\cite{Kobayashi:2019xdt}, with $\mu$ being the electron-to-proton mass ratio.
Limits on oscillations of the nuclear parameters with frequencies in the range of \SI{10}{\Hz} to \SI{100}{\mega\Hz} were obtained using molecular spectroscopy from vibrational transitions in the neutral iodine molecule, I$_2$~\cite{Oswald:2021vtc}.

A unique subset of molecules is the homonuclear diatomic molecules such as Sr${}_2$~\cite{Kondov:2019jzq}, O${}_2^+$~\cite{Carollo:2018arm,Wolf:2020sew,Hanneke:2020xsm}, or N${}_2^+$~\cite{kajita2014test}. The symmetry of these apolar molecules dictates that rotational and vibrational transitions of the same electronic state are electric-dipole forbidden \cite{tino2022identical, germann2014observation}. The lack of dipole coupling makes these molecules ideal for creating molecular qubits \cite{najafian2020megahertz} and clocks \cite{leung2023terahertz}, as their excited vibrational-state natural lifetimes can extend to months.

A direct comparison between a molecular clock and an optical clock will typically require a frequency comb to gap between the distinct transition frequencies of the two clocks. 
A class of methods to overcome the above problem are known as ``enhanced-sensitivity'' methods \cite{flambaum2007enhanced,demille2008enhanced,hanneke2016high,pavsteka2015search, Kozyryev:2018pcp}. These methods allow measuring an energy level with optical-frequency sensitivity in the microwave-frequency domain due to an accidental degeneracy in the spectrum. A few examples are the constraints put on drifts in the fine-structure constant and $\mu$ by exploiting degenerate transition in atomic Dy \cite{leefer2013new} and molecular KRb \cite{Kobayashi:2019xdt}, respectively.
Experiments to probe oscillations of $\mu$ with nearly-degenerate transitions between vibrational levels in different modes of polyatomic molecules, for instance, SrOH~\cite{Kozyryev:2018pcp}, have also been proposed.

In this letter, we propose an ``enhanced-sensitivity'' method that does not rely on an internal degenerate transition within the atom or the molecule. Instead, we propose to use correlation spectroscopy \cite{chwalla2007precision,chou2011quantum,clements2020lifetime, manovitz2019precision} between transitions of two different clocks that are naturally degenerate. Here, we consider transitions as degenerate if the difference in frequency is below $\Delta f \lesssim \SI{20}{\GHz}$. The remaining frequency gap between the clocks is bridged using an electro-optic modulator. In this case of degeneracy, the same laser can be used to interrogate both clocks simultaneously. This way, laser noise, which is a common-mode noise for both clocks, does not degrade the measurement sensitivity. We show that due to the dense spectrum of the heavy molecular iodine ion, I$_2^+$, a nuclear-parameters-sensitive transition that is degenerate (up to a microwave frequency) with the nuclear-parameters-insensitive clock transition in Ca$^+$ exists.

We calculate the absolute sensitivity of the molecular transition to changes in $\mu$ and estimate our scheme's spectroscopic stability. With that, we calculate the method's sensitivity for ULDM searches. 
Our projected sensitivity for both scalar and axion ULDM can extend beyond the reach of current laboratory tests, depending on the mass range.
For axion DM, however, constraints from density corrections to the axion potential, for instance, inside or on the surface of the Earth, preclude the parameter space in which our scheme yields the strongest constraints.
We particularly provide some analysis regarding the robustness of these constraints in the context of the \ZN QCD-axion model~\cite{Hook:2018jle}.
Our proposed method projects a few-orders-of-magnitude improvement in ULDM sensitivity over best atomic clock comparisons.

%
\section{Iodine molecular ion}
%

Pa\v{s}teka \etal~\cite{pavsteka2015search} suggested using dihalogen molecular ions for searching variations of fundamental constants by exploiting degenerate transitions between the two spin-orbit manifolds of their $X\,^2\Pi_\Omega$ electronic ground states. Here, we look at the homonuclear dihalogen molecular iodine ion, I$_2^+$, as a sensitive ULDM detector. The molecule's heavy mass of \SI{254}{\atomicmassunit} dictates a dense rovibrational spectrum. We propose to utilize this dense spectrum for correlation spectroscopy \cite{chwalla2007precision,chou2011quantum,clements2020lifetime, manovitz2019precision} with the Ca$^+$ optical atomic-ion clock by exploiting frequency-degeneracy between the molecular and atomic clocks' transitions.

\begin{figure*}
    \includegraphics[width=\linewidth]{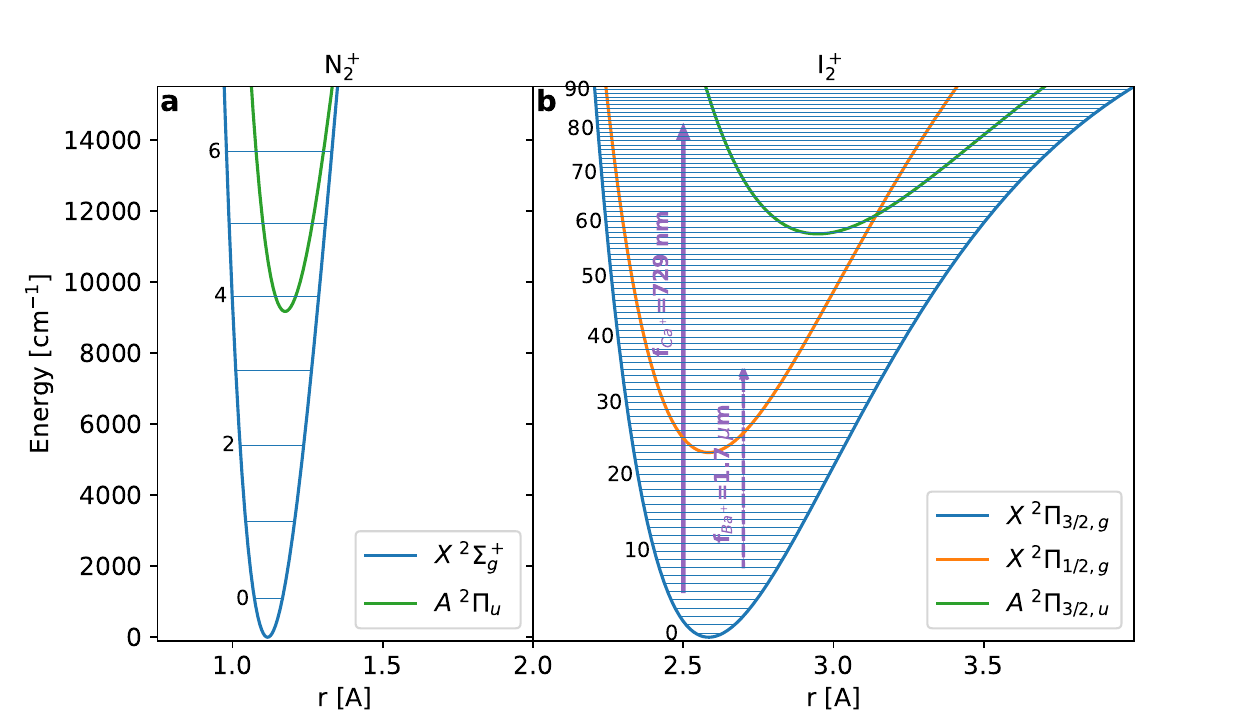}
    \caption{The vibration spectrum in the electronic ground state of N$_2^+$ (a) and I$_2^+$ (b) molecular ions (blue horizontal lines). Electronic ground (blue), spin-orbit (orange), and excited (green) states are calculated under the Morse potential approximation. Spectroscopic values are taken from Refs.~\cite{deng2012high,cockett1996zero,cockett1995zero,ferguson1992high}. The purple arrows correspond to the transition frequencies of Ca$^+$~\cite{BIPM} (solid line) and Ba$^+$~\cite{arnold2020precision} (dashed line, see discussion) optical atomic-ion clocks.  
    }
    \label{fig:spectrum}
\end{figure*}

The spectrum of the lowest electronic states of I$_2^+$, and for comparison, N$_2^+$, are given in \namecrefs{fig:spectrum}~\labelcref{fig:spectrum}(b) and~\labelcref{fig:spectrum}(a), respectively. The much denser spectrum of the heavy I$_2^+$ as compared to N$_2^+$ is visually striking. It opens up opportunities for new-physics searches as discussed in this paper.

\begin{figure}
    \includegraphics[width=\linewidth]{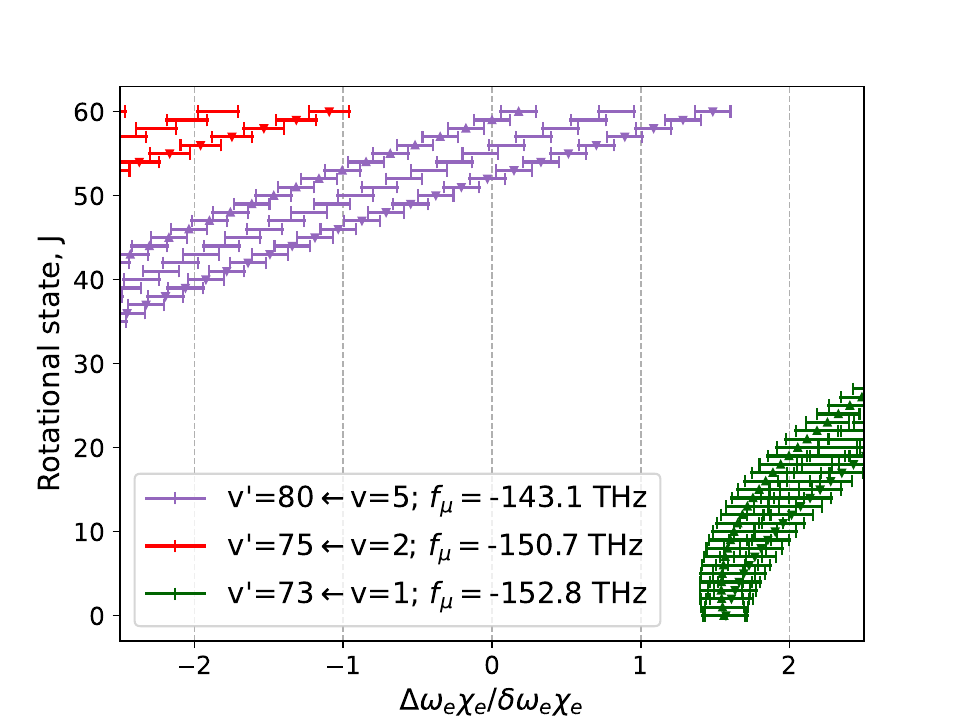}    
    \caption{Degeneracy of I$_2^+$ and Ca$^+$ clocks. 
    The dominant term for the uncertainty in the I$_2^+$ transition frequency is the uncertainty in the anharmonic-vibrational correction constant, $\delta \omega_e\chi_e$. The $x$-axis represents a possible deviation of $\omega_e\chi_e$ from its literature value, $\Delta \omega_e\chi_e$.
    The error bars represent the range of $\omega_e\chi_e$ values for which $|f_{\textrm{I}_2^+}-f_{\textrm{Ca}^+}|<\SI{20}{\GHz}$.  
    We indicate different vibrational transitions with different colors (see legend). The $y$-axis represents the ground-rotational state of the transition. We indicate a change in the rotational quantum number, $\Delta J=1,-1, 0$, by triangles pointing up, pointing down, or no triangles, respectively.}
    \label{fig:degeneracy}
\end{figure}

The rovibrational energy levels of the $X\,^2\Pi_{3/2,g}$ electronic ground state are given by \cite{deng2012high}
\begin{align}\begin{aligned}
    E_{v,J}&=\omega_e\left(v+\frac{1}{2}\right)-\omega_e\chi_e\left(v+\frac{1}{2}\right)^2\\
    &\phantom{=}\quad +\left(B_e-a_e\left(v+\frac{1}{2}\right)\right)J(J+1)\,.
\end{aligned}\end{align}
Here, $\omega_e$ is the vibrational constant, $\omega_e\chi_e$ is the anharmonic vibrational correction, $B_e$ is the rotational constant, and $a_e$ is its vibrational correction.
We use the experimentally measured spectroscopic values of Deng \etal~\cite{deng2012high} to calculate rovibrational transition frequencies (a unit conversation factor is implicit), $f_{v',J'\leftarrow v,J}=E_{v',J'}-E_{v,J}$, that match that of the clock transition of Ca$^+$, $f_{\textrm{Ca}^+}=\SI[separate-uncertainty=false]{411042129776400.4(7)}{\Hz}$~\cite{BIPM}.
We found that the transition in I$_2^+$, $f_{\textrm{I}_2^+}=f_{80,51\leftarrow 5,52}$, is degenerate with the Ca$^+$ clock transition to within the uncertainty of the known spectroscopic values, $\Delta f=f_{\textrm{I}_2^+}-f_{\textrm{Ca}^+}=\SI{5\pm169}{\GHz}$ (\cref{fig:spectrum}b purple solid arrow). The main contribution to this uncertainty stems from the anharmonic vibration term, $\delta(\Delta f)\approx\delta \omega_e\chi_e(v'-v)(v'+v+1)$ with $\delta \omega_e\chi_e\approx\SI{0.00087}{\per\cm}$~\cite{deng2012high}. We confirmed that for different values of $\omega_e\chi_e$ within the spectroscopic error, $\delta\omega_e\chi_e$, there exists a degenerate transition in the microwave domain between the molecule and the atomic clocks (\cref{fig:degeneracy}). This is possible due to the dense spectrum of the I$_2^+$ molecule that allows tuning the transition frequencies by choosing different vibrational and rotational quantum numbers.

The absolute sensitivity of a molecular transition to a change in the proton-to-electron mass ratio, $f_{\mu;v'\!,v} = \mu\, d f_{v'\!,v}/d \mu$, is given by
\begin{equation}\label{eq:abs_sens}
    f_{\mu;v'\!,v} = -\frac{1}{2}\omega_e(v'-v)\left(1-2\frac{\omega_e\chi_e}{\omega_e}(v'+v+1)\right).
\end{equation}
Here, we neglected the small contribution of the rotational constants to the absolute sensitivity and higher anharmonic vibrational terms. From \cref{eq:abs_sens}, we see that for increasing the sensitivity, we need to consider a large overtone transition, $v'-v\gg1$. In addition, for high vibration ($v'+v\gg1$), the anharmonic-vibration-correction constant, $\omega_e\chi_e$, tends to decrease the sensitivity. 
The absolute sensitivity of vibration transition compatible with correlation spectroscopy in our system are given in the legend of \cref{fig:degeneracy}.

%
\section{Correlation spectroscopy}
%

Correlation spectroscopy~\cite{chwalla2007precision,chou2011quantum,clements2020lifetime, manovitz2019precision} is a powerful method to increase the probe time beyond the laser coherence time up to the fundamental limit given by the transition's natural lifetime.
In the context of new-physics searches, we propose to use it to extract the difference in the beating of two clocks with different sensitivities to oscillating DM fields. 
The method circumvents the need for a frequency comb to compare the two clock's optical frequencies. Instead, the correlation-spectroscopy signal is already generated in the microwave regime, where it can be referenced to a commercial microwave clock. 
The method also slightly relaxes the need for ultra-stable and sub-Hz laser sources for clock interrogation as the laser noise is a common-mode noise that factors out in this method.

Denoting the ground and excited state of clock~$i$ ($i=1,2$) by $|g_i\rangle$ and $|e_i\rangle$, respectively, the state of the two clocks during the Ramsey correlated interrogation is given by
\begin{align}\begin{aligned}
    |\psi\rangle &= 
    \frac{|g_1\rangle +e^{i(\varphi_1+\varphi_N)}|e_1\rangle}{\sqrt{2}}
    \otimes
    \frac{|g_2\rangle +e^{i(\varphi_2+\varphi_N)}|e_2\rangle}{\sqrt{2}}\\ 
    &= \begin{multlined}[t]\frac{1}{2}\big[|g_1 g_2\rangle + e^{i(\varphi_1+\varphi_2+2\varphi_N)}|e_1 e_2\rangle \\ 
    +e^{i\varphi_N}(e^{i\varphi_1}|e_1 g_2\rangle+e^{i\varphi_2}|g_1 e_2\rangle)\big]. \end{multlined}
\end{aligned}\end{align}
Here, $\varphi_i=\Delta_i T_R + \tilde\varphi_i$ is a deterministic phase accumulated during the Ramsey interrogation time, $T_R$, where $\Delta_i$ is laser's detuning from resonance, and $\tilde\varphi_i$ is a user-controlled phase. The laser noise, $\varphi_N$, is common for both clocks and results in fast decoherence of half of the signal while the other half is immune to it.  

The correlation signal is given by a parity measurement, $\hat\Pi=\hat\sigma_{z,1}\otimes\hat\sigma_{z,2}$, on the two clocks \cite{clements2020lifetime},
\begin{equation}
    \langle\hat\Pi\rangle=\frac{1}{2}\cos(\Delta_{21}T_R+\tilde\varphi_{21})e^{-\Gamma T_R}.
\end{equation}
Here, $\Delta_{21}=\Delta_{2}-\Delta_1$ and $\tilde\varphi_{21}=\tilde\varphi_2-\tilde\varphi_1$, and $\Gamma$ is the inverse of the excited-state natural lifetime. The factor 1/2 is due to the decoherence of half of the signal and the factor $e^{-\Gamma T_R}$ takes into account the finite lifetime of the clock's excited state.  

The fractional instability of the correlation-spectroscopy signal limited by quantum-projection noise and excited-state lifetime is given by \cite{clements2020lifetime}
\begin{equation}
    \sigma_c(\tau)=\frac{2}{2\pi f\sqrt{T_R \tau}}e^{\Gamma T_R},
\end{equation}
where $f$ is the transition frequency, 
and $\tau$ the total interrogation time. The minimal instability is achieved at a Ramsey pulse time that equals half the spontaneous lifetime, $T_R=\Gamma^{-1}/2$ \cite{clements2020lifetime}.

Since the lifetime of excited vibrational states of homonuclear molecules is estimated in months, the clock comparison instability will be limited by the excited-state lifetime of the Ca$^+$ clock, $\Gamma(\text{Ca}^+)=(\SI{1.16}{\s})^{-1}$~\cite{UDportal}. We choose a Ramsey pulse time of $T_R=\SI{0.58}{\s}$ and calculate the clock-comparison instability to be $\sigma_c(\tau)= \num{1.68e-15}/\sqrt{\tau/\si{\s}}$.

We evaluate the sensitivity for proton-to-electron mass ratio changes using our calculated transition absolute sensitivity,
\begin{equation}
    \frac{\delta \mu(\tau)}{\mu}=\frac{\delta f(\tau)}{f_\mu}=\sigma_c(\tau)\frac{f}{f_\mu}.
\end{equation}
Using the Ca$^+$-lifetime-limited instability of our system, the Ca$^+$ clock frequency, $f\approx\SI{411}{\THz}$, and the iodine's absolute sensitivity, $f_\mu\approx\SI{150}{\THz}$, we get $\delta \mu(\tau)/\mu\approx\num{4.6e-15}/\sqrt{\tau/\si{\s}}$. 

After some averaging time, $\tau_p$, the fractional instability of the clock will stop improving as $\sim1/\sqrt{\tau}$. This is the clock's precision, where uncontrolled fluctuations in the experiment parameters lead to oscillations in the clock signal not emanating from DM. We take a value of $\sigma_c\approx\num{e-18}$ as our clock-comparison precision. With this value, a sensitivity of $\delta\mu/\mu\approx\num{2.7e-18}$ is reached after averaging for $\tau_p\approx33$ days.

The clock's precision does not necessarily limit the clock-comparison sensitivity. For example, astrophysical observations with limited precision of $\delta \mu/\mu\sim\num{e-7}$ can limit drifts of $\SI{e-17}{\per\year}$ assuming a linear variation of $\mu$ over a coarse of $\SI{e10}{\year}$~\cite{Safronova:2017xyt}.
The maximum sensitivity to drifts in $\mu$ assuming continuous measurement over a period of time, $\tau$, is given by
\begin{equation}
    \partial_t \log\mu=2\sqrt{2}\, \frac{\delta \mu(\tau)}{\mu}\,\tau^{-1}.
\end{equation}
For our system parameters, the drift sensitivity is $\partial_t \log\mu\approx\num{1.2e-14}/(\tau/\si{s})^{3/2}\; [\si{\per\s}]$. After averaging for 31~hours, we can reach a drift sensitivity of $\SI{e-14}{\per\year}$ as obtained in the KRb experiment~\cite{Kobayashi:2019xdt}. A one-year campaign will result in drift sensitivity of $\SI{2.3e-18}{\per\year}$, an order of magnitude better sensitivity than the most stringent bound set to date on drifts~\cite{lange2021improved}.

%
\section{ULDM sensitivity}
%

We consider here two generic spin-zero ULDM candidates to interpret our results as limits on the DM coupling: a scalar field~$\phi$ with mass~$m_\phi$ 
and an axion field~$a$ of mass~$m_a$ with pseudoscalar couplings to QCD.

The interactions of the scalar field with the QCD part of the SM at low energies are governed by the Lagrangian
\begin{equation}
    \label{eq:dilatonLag}
    \mathcal{L} \supset -\kappa \phi\,  \frac{d_g \beta_s}{2\,g_s}\,G^a_{\mu\nu} G^{a\,\mu\nu}.
\end{equation}
Here, $g_s$ and $\beta_s$ are the QCD coupling constant and $\beta$-function, respectively, 
$G_{\mu\nu}^a$ is the gluon field-strength tensor, and $\kappa = \sqrt{4 \pi G_N}$ with the Newton constant~$G_N$.
The normalization of the dilaton coupling~$d_g$ is chosen such that the couplings are renormalization group invariant~\cite{Damour:2010rm,Damour:2010rp}.
For the axion, the coupling to gluons is given by
\begin{align}
    \label{eq:axionLag}
    \mathcal{L} &\supset \frac{g_s^2}{32 \pi^2}\,\frac{a}{f_a}\, \tilde{G}^a_{\mu\nu} G^{a\,\mu\nu} \,,
\end{align}
where $\tilde{G}^a_{\mu\nu} = \frac{1}{2} \epsilon_{\mu\nu\alpha\beta} G^{a\,\alpha\beta}$ is the dual gluon field-strength and $f_a$ is the axion decay constant.

The scalar interactions in~\cref{eq:dilatonLag} induce oscillations in the QCD scale,~\LQCD, 
with a corresponding fractional change linear in the scalar field,
\begin{equation}
    \frac{\Delta\LQCD}{\LQCD} = d_g \kappa\phi\,.
    \label{eq:dilatonFC}
\end{equation}
The axion interaction~\cref{eq:axionLag}, on the other hand, causes oscillations of the pion mass~$m_\pi$, 
\begin{align}
    m_\pi^2(\theta) = m_\pi^2\ \sqrt{1 - \frac{4\, m_u m_d}{(m_u+m_d)^2} \sin^2\frac{\theta}{2}}\,,
    \label{eq:pion_mass}
\end{align}
where $\theta = {a}/{f_a}$, and $m_u$ and $m_d$ are up and down quark masses. To leading order, the oscillations are quadratic in the DM field,
\begin{align}
    \frac{\Delta m_\pi^2}{m_\pi^2} =   - \frac{m_u m_d}{2 (m_u + m_d)^2}\,\theta^2 \,.
    \label{eq:axionFC}
\end{align} 
The changes in \LQCD and $m_\pi^2$ then, in turn, lead to oscillations in the nucleon mass,~$m_N$, and hence in the electron-to-proton mass ratio, related to \cref{eq:dilatonFC,eq:axionFC} by
\begin{equation}
    \frac{\Delta m_N}{m_N} \sim \frac{\Delta\LQCD}{\LQCD}\sim \kappa d_g \frac{\sqrt{2\rho_{\DM}}}{m_\phi}\cos(m_\phi t),
\end{equation}
and ~\cite{Kim:2022ype}
\begin{equation}
    \label{eq:delta_mp_axion}
    \frac{\Delta m_N}{m_N} \sim \frac{\sigma_{\pi N}}{m_N}\, \frac{\Delta m_\pi^2}{m_\pi^2}\sim 0.006 \frac{\rho_{\DM}}{m_a^2\,f_a^2}\cos(2 m_a t),
\end{equation}
respectively, where $\sigma_{\pi N} \sim \SI{60}{\MeV}$~\cite{Alarcon:2011zs} is the nuclear-$\sigma$ term.%
\footnote{Results for $\sigma_{\pi N}$ from lattice simulations range from \SIrange{30}{80}{\MeV}~\cite{FlavourLatticeAveragingGroupFLAG:2021npn}.}
Note that the quadratic coupling of the axion field leads to oscillations in twice the Compton frequency.

%
\section{ULDM searches}
%

\begin{figure*}
    \includegraphics{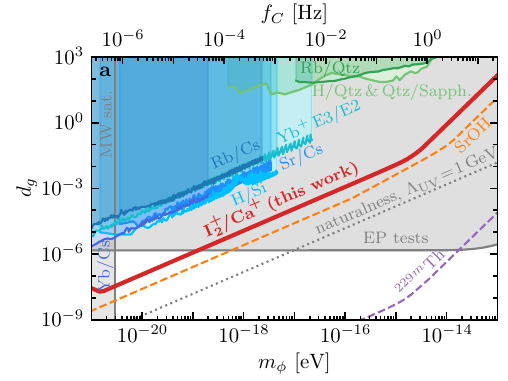}
    \hfill
    \includegraphics{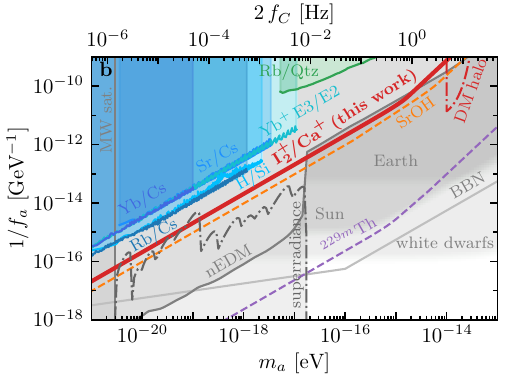}
    \caption{%
        Constraints on the scalar coupling $d_g$ to gluons~(a) and axion decay constant $f_a$~(b) as a function of the DM mass. 
        The thick red line indicates the projected constraint from correlation spectroscopy derived in this work.
        Current limits from atomic spectroscopy are indicated by solid colored lines, whereas the projections for future experiments are shown as dashed lines. 
        Non-spectroscopic bounds are shown in gray (see more detailed discussion in the text).
        The region below the dotted line in (a) is motivated by naturalness arguments for the scalar mass if a cutoff around \SI{1}{\GeV} is assumed.
        The red dash-dotted line in (b) indicates the sensitivity enhancement in the presence of an axion-DM halo around the Sun.
    }
    \label{fig:results}
\end{figure*}

In \cref{fig:results}, we show the projected sensitivity for the scalar~[\cref{fig:results}(a)] and axion~[\cref{fig:results}(b)] coupling to gluons from correlation spectroscopy of I$_2^+$ and Ca$^+$~(thick red line) as a function of the DM mass.
We account for stochastic fluctuations of the DM amplitude, leading to a reduction of the limits for measurements shorter than the coherence time by a factor of $\sim 3$~\cite{Centers:2019dyn}.
We assume that measurements are taken at a rate determined by the Ramsey time over the course of one month.
We are hence primarily sensitive to DM masses between $2\pi/(\SI{31}{\day})^{-1} \approx \SI{e-21}{\eV}$ and $2\pi/(\SI{.58}{\second})^{-1} \approx \SI{e-14}{\eV}$, where our setup can probe scalar couplings of $d_g \gtrsim \num{e-5}\,\frac{m_\phi}{\SI{e-18}{\eV}}$ and decay constants around $f_a \lesssim \SI{2e14}{\GeV}\,\frac{\SI{e-18}{\eV}}{m_a}$.

DM fields with higher masses can still be probed via aliasing~\cite{Derevianko:2016vpm} or dynamical decoupling~\cite{Aharony:2019iad}, while DM oscillating at lower frequencies produces a drift-like signal.
The observations of Milky Way satellite galaxies, however, impose a lower bound on the ULDM mass of $m_\DM \gtrsim \SI{2.9e-21}{\eV}$~\cite{DES:2020fxi}, indicated by the vertical gray lines in \cref{fig:results}.
For axion masses around $m_a\sim\SI{e-14}{\eV}$, the local DM density can further be enhanced as the Sun can capture the axion, forming a solar DM halo~\cite{Budker:2023sex}.
As the DM induced changes in the nucleon mass scale with the DM density, cf.\ \cref{eq:delta_mp_axion}, this also amends the constraints from clock experiments.
The corresponding amelioration of our projected sensitivity is depicted by the red dashed-dotted line in \cref{fig:results}(b).

For comparison, we also depict the constraints on gluon-coupled-ULDM oscillations from existing clock experiments as solid colored lines.
Atomic clocks~(blue) involving hyperfine transitions in comparisons of Rb/Cs~\cite{Hees:2016gop}, H/Si~\cite{Kennedy:2020bac},  Yb/Cs~\cite{Kobayashi:2022vsf} and Sr/Cs~\cite{Sherrill:2023zah} are sensitive to DM masses below \SI{e-17}{\eV} and couplings around $d_g \gtrsim \num{e-3}\,\frac{m_\phi}{\SI{e-18}{\eV}}$ and $f_a \lesssim \SI{e13}{\GeV}\,\frac{\SI{e-18}{\eV}}{m_a}$, respectively.
A similar mass and coupling range can be constrained from DM-induced variations of the charge radius in octupole~(E3) and quadrupole~(E2) electric transitions in an Yb${}^+$ atomic clock~\cite{Banerjee:2023bjc}~(cyan).
Acoustic oscillators~(green) based on H/Quartz-Quartz/Sapphire~\cite{Campbell:2020fvq} and Rb/Quartz~\cite{Zhang:2022ewz} can reach up to $m_\DM \sim \SI{e-12}{\eV}$ with $d_g \gtrsim \numrange{10}{100}$ and $f_a\lesssim\SI{e10}{\GeV}$ at the lower mass end (around \SI{e-18}{\eV}), whereas molecular spectrosopy of neutral~I$_2$ (not shown in the plot) can probe $d_g\gtrsim\num{e5}$ and $f_a \lesssim \SI{e6}{\GeV}$ for masses above \SI{e-14}{\eV}~\cite{Oswald:2021vtc}.
The prospective reach of our proposed experiment hence exceeds the sensitivity of current clock probes of oscillating DM by orders of magnitude.

We further indicate the projected sensitivities of a molecular SrOH clock~\cite{Kozyryev:2018pcp}~(orange) and a hypothetical nuclear clock based on ${}^{229m}$Th~\cite{Arvanitaki:2014faa,Banerjee:2020kww,Peik:2020cwm}~(purple) by dashed lines, assuming a one-year campaign length in both.
While the projection for the former is comparable to what can be achieved with our I${}_2^+$/Ca${}^+$ setup within half an order of magnitude%
\footnote{In Ref.~\cite{Gue:2024onx}, similar bounds have been projected for atom interferometry constraints from the MICROSCOPE mission~\cite{Touboul:2017grn}.} 
(as both proposals rely on molecular vibration transitions), the nuclear clock exhibits an improvement of another six orders of magnitude. This is due to a degeneracy that allows measuring a $\sim$MeV-sensitive nuclear-transition energy at the $\sim$eV scale.
However, the technological developments of nuclear clocks are still in their infancy compared to molecular clocks.

Light scalars are further subject to constraints from fifth force experiments and equivalence principle~(EP) tests~\cite{Damour:2010rp,Hees:2018fpg}.
As these bounds are based on the corresponding Yukawa force mediated between free-falling objects, they are independent of the cosmological abundance, i.e., they do not depend on whether the scalar constitutes DM or not.
For comparison, we depict the measurement results of the E\"{o}tv\"{o}s parameter by the Eot-Wash group~\cite{Smith:1999cr,Schlamminger:2007ht} and the MICROSCOPE experiment~\cite{MICROSCOPE:2022doy}. The corresponding upper bound on the scalar coupling of $d_g \lesssim \num{e-6}$ is shown in gray in \cref{fig:results}(a).

From the theoretical side, the scalar couplings and masses can be constrained by naturalness arguments.
The coupling to gluons induces corrections to the scalar mass on the order of $\Delta m_\phi^2 \sim\frac{d_g^2 G_N \Lambda_\mathrm{UV}^4}{16 \pi^2}$, where $\Lambda_\mathrm{UV}$ is the UV cutoff scale.
The naturalness bound is obtained by requiring that these mass corrections do not exceed the mass itself, $\Delta m_\phi^2 \lesssim m_\phi^2$.
The corresponding couplings are indicated by the gray dotted line in \cref{fig:results}a, assuming a cutoff of $\Lambda_\mathrm{UV} \simeq \SI{1}{\GeV}$, where the region below the line is motivated by naturalness.

Additional constraints on the axion-gluon coupling arise from various couplings to nuclei generated by the interaction of \cref{eq:axionLag}. 
The strongest bounds at low axion masses are obtained from searches for an oscillating neutron electric dipole moment~(nEDM)~\cite{Abel:2017rtm,Roussy:2020ily,Schulthess:2022pbp}, constraining $f_a \gtrsim \SI{e16}{\GeV}\,\frac{\SI{e-18}{\eV}}{m_a}$ for $m_a \lesssim \SI{e-17}{\eV}$ as indicated in gray in \cref{fig:results}b.
From astrophysics, the axion nuclear coupling is constrained to $f_a \gtrsim \SI{4e8}{\GeV}$ by supernova~(SN)~\cite{Raffelt:2006cw,Carenza:2019pxu,Caputo:2024oqc} (see, however, Ref.~\cite{Bar:2019ifz}) and neutron star~(NS) cooling bounds~\cite{Leinson:2021ety}.
Furthermore, an axion-DM-induced increase of the neutron-proton mass difference during Big Bang nucleosynthesis~(BBN) would reduce the ${}^4$He mass abundance~\cite{Blum:2014vsa}, excluding $f_a \lesssim \SI{e16}{\GeV} \min\left[\left(\frac{m_a}{\SI{e-16}{\eV}}\right)^{-1},\left(\frac{m_a}{\SI{e-16}{\eV}}\right)^{-1/4}\right]$ as indicated by the light gray line.
This bound, however, can be avoided, e.g.,\ through a non-standard cosmological evolution of the axion, such that the values of $\theta$ at BBN and today are not related via the usual relation.

In minimal QCD axion models, the interaction in \cref{eq:axionLag} induces a potential for the axion given by 
$V(\theta) = - f_\pi^2\, m_\pi^2(\theta)$, 
where $f_\pi \simeq \SI{92}{\MeV}$ is the pion decay constant.
This establishes a relation between the axion's mass and its decay constant,
$m_a = \frac{\sqrt{m_u m_d}}{m_u+m_d}\,\frac{m_\pi f_\pi}{f_a} \simeq \SI{e-5}{\eV}\,\frac{\SI{e12}{\GeV}}{f_a}$~\cite{Weinberg:1977ma,GrillidiCortona:2015jxo}, the so-called QCD axion line.
Clock experiments, similar to most other current experiments searching for ultralight axions, struggle to reach the corresponding couplings required to obtain low axion masses, requiring the consideration of models that do not follow this relation.
Masses significantly below the mass predicted by the QCD axion line can be realized in technically natural models without fine-tuning.
For instance, in a \ZN symmetric model consisting of \Ncopy identical copies of the SM, each with a $\theta$ parameter shifted by $2\pi/\Ncopy$ with respect to the neighboring sectors and interacting only through a common axion field, the axion mass resulting from the summation over the \Ncopy sectors is suppressed exponentially by a factor $(m_u/m_d)^\Ncopy$ compared to the canonical QCD axion mass~\cite{Hook:2018jle}.

Superradiant axion production in the vicinity of spinning black holes~\cite{Arvanitaki:2014wva}, on the other hand, puts upper limits on the DM self-interaction.
\Cref{fig:results}(b) shows superradiance bounds from supermassive black holes~\cite{Unal:2020jiy} as a dash-dotted gray line, assuming self-interactions $\lambda \sim m_a^2/f_a^2$, corresponding to the canonical cosine potential of the axion. 
This constraint however changes if the self-interactions are non-standard, and is, for instance, in the \ZN model for light axions suppressed by the number of copies.

For axions lighter than the canonical QCD axion, density corrections can flip the sign of the axion potential inside compact objects such as Earth.
If the Compton wavelength of the axion inside Earth is smaller than the Earth's radius, the axion field shifts to $\theta=\pi$ in the interior, which leads to non-zero $\theta$ on the surface.
This constrains the decay constant for light axions to $f_a \gtrsim \sqrt{\rho_\oplus} R_\oplus \sim \SI{e13}{\GeV}$, where $\rho_\oplus$ and $R_\oplus$ are the Earth's density and radius, respectively~\cite{Hook:2017psm}.
The corresponding axion masses and couplings at which the axion develops a static profile around Earth are indicated by the gray-shaded region in \cref{fig:results}(b). 
Stronger bounds around $f_a \gtrsim \SI{e16}{\GeV}$ can be obtained from the same effect considering processes in the solar core or the mass-radius relation of white dwarfs~\cite{Balkin:2022qer}.
In addition, it can induce a new ground state of nuclear matter at finite density, extending the mass range of neutron stars beyond the QCD prediction~\cite{Balkin:2023xtr}.

%
\section{A comment on the Earth bound}
%

Before moving forward, let us comment on the robustness of the Earth, Sun, and white dwarfs density-correction bounds in \cref{fig:results}(b). In the following we will focus on the Earth bound, even though similar arguments can be made on all density-correction bounds. 
Unlike the bounds arising from direct ULDM searches, it is inferred from the fact that the density of the Earth changes the structure of the axion potential~\cite{Hook:2018jle}.
Similar to our molecular clock bound, these density corrections stem from the $\theta$-dependence of the nucleon mass.
This in turn induces density corrections to the axion mass, 
\begin{align}
    \label{eq:ma_density_corrections}
    \Delta m_a^2(\rho_N) = \frac{\rho_N}{f_a^2} \frac{\partial m_N}{\partial m_\pi^2} \frac{\partial^2 m_\pi^2}{\partial \theta^2} = \frac{\sigma_{\pi N} \,\rho_N}{f_a^2} \frac{1}{m_\pi^2} \frac{\partial^2 m_\pi^2}{\partial \theta^2} ,
\end{align}
where $\rho_N$ is the nucleon number density and $\frac{\partial^2 m_\pi^2}{\partial \theta^2} < 0$ at the origin, cf.\ \cref{eq:axionFC}.
In a dense medium, as, for instance, inside Earth, the minimum of the axion potential at $\theta=0$ can hence turn into a maximum, giving rise to the bound in \cref{fig:results}(b).

The Earth bound and our molecular clock limits are therefore interrelated.
As a consequence, an alleviation of the former in general also leads to a mitigation of the latter, provided that the sign of the density corrections in \cref{eq:ma_density_corrections} is unchanged.
Going to the extreme case, if additional contributions to the nucleon mass render the density correction $\Delta m_a^2(\rho_N)$ positive, the Earth bound is completely removed. 
In this case, the origin remains a minimum of the axion potential, even in a dense medium.
Clock experiments, on the other hand, do not depend on the sign of the axion-nucleon coupling and would, therefore, still be sensitive.
Such a drastic change of the axion interaction is, however, non-trivial to achieve.
In particular, to comply with constraints from nuclear physics, the pion-nucleon interaction should not be modified significantly.
Furthermore, care has to be taken to prevent the phases of additional couplings from spoiling the solution of the strong $CP$ problem.

To this end, first, consider the nuclear mass term in two-flavor chiral perturbation theory. 
The leading mass correction comes from the term 
$\mathcal{L} \supset c_1 \langle\chi_+\rangle \bar{N}N$~\cite{Gasser:1987rb,Fettes:2000gb,Scherer:2012xha},
where $c_1 = -\SI{1.1}{\GeV^{-1}}$~\cite{Hoferichter:2015tha}
is a low-energy constant, $N=(p,n)$ is the nucleon isospin doublet, and $\langle\chi_+\rangle = \frac{m_\pi^2 f_\pi^2}{2\,(m_u+m_d)} \tr\left[\mathcal{M}_a\, U^\dagger + \hc\right]$.
Here, $\mathcal{M}_a$ is the axion-dependent quark mass matrix, i.e., after the $\theta$-term, \cref{eq:axionLag}, has been removed via a chiral rotation of the quark fields,
and $U = \exp(i\,\Pi^a \sigma^a/f_\pi)$ with $\sigma^a$ being the Pauli matrices describes the pion fields~$\Pi^a$.
This gives the tree-level contribution to the $\theta$-dependence of the nucleon mass.
At one-loop, the latter is given by~\cite{Gasser:1987rb,Scherer:2012xha}, 
\begin{align}
    \label{eq:nucleon_mass}
    m_N(\theta) = m_0 - 4\,c_1\,m_\pi^2(\theta)  - \frac{3\,g_A^2\,m_\pi^3(\theta)}{32 \pi\,f_\pi^2}\,,
\end{align}
where $m_0$ is the nucleon mass in the chiral limit ($m_u,m_d \to 0$), $m_\pi^2(\theta)$ is the axion-dependent pion mass in \cref{eq:pion_mass}, and $g_A=1.27$~\cite{ParticleDataGroup:2022pth} is the nucleon axial-vector coupling to pions.

Our main idea is to modify the model such that an additional contribution to the nuclear mass is induced, 
\begin{align}
    \label{eq:Delta_mN_NP}
    \Delta m_N = -4\,\tilde{c}\,m_\pi^2(\theta)\,.
\end{align}
We do so in the context of the \ZN axion model~\cite{Hook:2018jle}, consisting of \Ncopy copies of the SM with a \ZN symmetry and interacting only through the axion.
We extend the model by additional couplings between the nucleons of each copy~$(i)$ and the pions of the neighboring sectors~$(i\pm1)$,
\begin{align}
    \label{eq:LagNP}
    \Delta \mathcal{L} \supset \sum\limits_{i=1}^\Ncopy \frac{1}{2}\,\tilde{c} \, \left(\langle \chi_+^{(i+1)}\rangle + \langle \chi_+^{(i-1)}\rangle\right) \bar{N}^{(i)}N^{(i)} \,.
\end{align}
In the limit of large \Ncopy, we obtain the nucleon mass correction in \cref{eq:Delta_mN_NP}.
Alternatively, \cref{eq:Delta_mN_NP} might be generated via interactions with an additional scalar with density-dependent expectation value, as, e.g.,\ introduced in a different context to circumvent stellar cooling bounds in Refs.~\cite{Budnik:2020nwz,Bloch:2020uzh,DeRocco:2020xdt}.

With \cref{eq:Delta_mN_NP}, the density correction to the axion mass is modified to 
$\Delta m_a^2  = \frac{m_u m_d}{(m_u+m_d)^2} (4 \,\tilde{c}\,m_\pi^2 - \sigma_{\pi N})\, \rho_N/f_a^2$.
The coupling probed by our I$_2^+$/Ca$^+$ clock comparison and the Earth bound are obtained replacing $1/f_a$ in \cref{eq:delta_mp_axion} by 
\begin{align}
    \label{eq:fa_rescaling}
    \frac{1}{f_a} 
    \quad\longrightarrow\quad
    \frac{1}{f_\text{eff}} = \frac{1}{f_a} \ \sqrt{\left|\frac{4\, \tilde{c}\, m_\pi^2}{\sigma_{\pi N}}-1\right|}\,.
\end{align}
Note that other clock comparisons involving hyperfine transitions also depend on the gyromagnetic $g$-factor of the nucleus~\cite{Kim:2022ype}, and hence scale differently compared to \cref{eq:fa_rescaling}. 
In the Rb/Cs clock comparison, for instance, the direct dependence on the nuclear mass is canceled to leading order, so that the bound is not affected by the additional interaction.

Completely removing the Earth bound requires $\tilde{c} \gtrsim  \tilde{c}_\mathrm{crit} = \frac{\sigma_{\pi N}}{4\, m_\pi^2} \approx \SI{0.81}{\per\GeV}$.
The applicability of an effective theory with such a low cutoff is certainly questionable.
Nonetheless, the construction in \cref{eq:LagNP} might still be of interest in the context of other experiments probing different axion couplings with a sensitivity close to the Earth bound.
However, already for $\tilde{c} \sim \SI{0.1}{\per\GeV}$, the bound is relaxed by less than \SI{10}{\%}.
For illustration, we still consider $\tilde{c}\gtrsim \tilde{c}_\mathrm{crit}$ in the following.

A further investigation of whether the model can escape other constraints is deferred to future work.
In particular, as we introduce couplings between nucleons and dark pions of similar size as the corresponding coupling within the SM, \cref{eq:LagNP} is subject to presumably strong constraints from missing energy searches.
Note, however, that since \cref{eq:LagNP} does not change the interactions between the SM pions and SM nucleons, QCD predictions are not altered significantly.
In addition, assuming a reheating temperature of a few \si{\MeV}, this term will not change the thermal history as in all sectors the pions acquire masses of the order of $\Lambda_\text{QCD}$ and the additional sectors, thus, do not thermalize with the SM.

In order to preserve the axion as a solution to the strong $CP$ problem, the phase of the coupling in \cref{eq:LagNP} further needs to be aligned with the QCD $\bar{\theta}$ parameter.
As the phase is a technically natural parameter, aligning the phases ad hoc is stable against radiative corrections.
Furthermore, the alignment could for instance be achieved by adding \Ncopy real scalar fields~$\phi^{(i)}$, which acquire a vacuum expectation value $\langle \phi^{(i)}\rangle = v_\phi$ and generate the light quark masses by effective interactions of the form
$\mathcal{L}^{(i)} \supset -\left[\frac{\phi^{(i)}}{v_\phi}\,\bar{q}^{(i)}_L\mathcal{M}_a^{(i)} q_R^{(i)} + \hc\right]$.\Cref{eq:LagNP} can then be generated by couplings to the gluons of the neighboring sectors, $\frac{\phi^{(i)}}{2\,\Lambda}\left[G^{a}_{\mu\nu} G^{a\,\mu\nu}\right]^{(i\pm1)}$ with $\Lambda > v_\phi$, after integrating out the scalars.
Assuming $m_\phi \sim v_\phi$, we obtain $\tilde{c} \sim \frac{f_\pi^2 m_N}{v_\phi^3 \Lambda}$, where we need to impose $v_\phi \gg \sqrt{m_\phi f_\pi} \sim \SI{100}{\MeV}$ to avoid corrections to the pion mass.
To reach $\tilde{c} = \tilde{c}_\mathrm{crit}$, we would however need $\Lambda v_\phi^3 \sim (\SI{220}{\MeV})^4$, pushing the effective theory beyond its limits.
The construction above can hence only be used to reduce the Earth bound by a factor of few at best.
While the construction of a more elaborate realization of \cref{eq:LagNP} and a proper exploration of the constraints on this scenario is beyond the scope of this letter, we can still discuss its implications on the Earth bound in the following.

\begin{figure}
    \centering
    \includegraphics{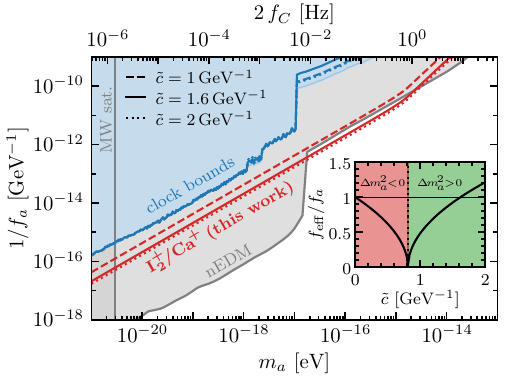}
    \caption{%
        Axion DM bounds from clock comparisons in the presence of the additional interaction in \cref{eq:LagNP} for different values of $\tilde{c} > \tilde{c}_\mathrm{crit}$.
        The inset depicts the factor by which the axion-nuclear coupling is modified as a function of $\tilde{c}$, with the Earth bound vanishing in the green region.
        For $\tilde{c}=\SI{1.6}{\per\GeV}$, the solid red line in the figure coincides with the one without the additional interaction shown in \cref{fig:results}b, i.e., $f_\text{eff}=f_a$. 
        Note that the nEDM bound does not depend on~$\tilde{c}$.
    }
    \label{fig:axion_unhooked}
\end{figure}

\Cref{fig:axion_unhooked} illustrates the dependence of the projected exclusion limits in this work and from existing clock experiments on the coupling $\tilde{c}$ in the presence of the additional term in \cref{eq:Delta_mN_NP}.
The inset depicts the rescaling factor $f_\text{eff}/f_a$, with the vertical dotted line indicating the critical value $\tilde{c}_\mathrm{crit}$. In the red region below $\tilde{c}_\mathrm{crit}$, both the density-effect bounds and our clock comparison limits are rescaled by the same factor, whereas in the green region above $\tilde{c}_\mathrm{crit}$, the Earth, Sun and white dwarf bounds are removed while the clock-comparison limits prevail. 
For completeness, we explore also sub-GeV values of $\tilde{c}^{-1}$, and find that for $\tilde{c}\simeq\SI{1.6}{\per\GeV}$, the rescaling factor is $f_\text{eff}/f_a=1$, i.e., the sensitivity of our clock experiment is the same as without the additional coupling.
Obviously, the sensitivity is enhanced for larger values of $\tilde{c}$, whereas it is reduced at lower values.
Note that the nuclear dipole moment is not sensitive to the coupling in \cref{eq:LagNP}, but rather directly to the value of $f_a$, so that, to leading order, the nEDM bounds do not depend on $\tilde{c}$.
The same applies for the superradiance bound, which we, however, omit in \cref{fig:axion_unhooked} as it probes the axion self-coupling instead of the coupling to nuclei.

We find that significant modifications of the bounds from density corrections to the axion potential are rather difficult to achieve with effective operators, even when neglecting potential constraints and taking the cutoff scale beyond the range of applicability of the effective theory.

%
\section{Experimental realization}
%

This section gives a concise description of our envisioned experimental realization. A single I$_2^+$ molecular ion will be trapped together with a single atomic ion in a radio-frequency ion trap in what is known as a quantum-logic configuration \cite{wolf2016non,chou2017preparation,sinhal2020quantum}. The atomic ion will be used for high-fidelity and quantum nondemolition state detection of the molecular-ion state \cite{sinhal2020quantum} by utilizing the shared motion of the two species in the trap. Due to the large mass of \SI{254}{\atomicmassunit} of $^{127}$I$_2^+$, we will use a heavy atomic ion, e.g., $^{176}$Yb$^+$ or $^{138}$Ba$^+$, for efficient motional coupling between the atomic and molecular ions. 

The molecular ion will be loaded into the trap from a pulsed molecular beam. A small amount of solid iodine will be heated to $\sim$\SI{80}{\celsius} to increase its vapor pressure. An inert seeding gas will be used to transport the I$_2$ molecules into the vacuum chamber and the ion trap \cite{shepperson2017strongly}. The state-selective resonance-enhanced multi-photon ionization method \cite{tong2012state} will be used to ionize the neutral iodine molecules into the I$_2^+$ rovibrational ground state \cite{cockett1995zero}. The rotational splitting in the ground state, $2B_e\approx$~\SI{2.4}{\GHz}, is large enough to be resolved with pulsed (few \si{\nano\second} long) lasers. Subsequent to their ionization, the molecules will be trapped in the deep potential created by the ion-trap's oscillating electric fields. Collisions with the laser-cooled atomic ions in the trap will lead to sympathetic cooling.

The ion trap will be cooled to~$\sim$\SI{4}{\kelvin} to improve the vacuum conditions and reduce the rate of state-changing and chemical-reaction collisions. These collisions could reduce our experiment's overall duty cycle if they exceed the rate at which we load a new molecule into the trap. Typical loading time should be between a few seconds to a few minutes. With the highly-reactive, Ar$^{13+}$, highly charged ion, a cryogenic ion-trapping setup achieved \SI{45}{\min} of trapping before a reaction occurred \cite{micke2020coherent}. We expect longer lifetimes for the I$_2^+$ molecule in the same vacuum conditions.

Correlation spectroscopy will be performed with a single laser locked to an ultra-stable low-expansion cavity. The laser will be split into two arms, each reaching a different clock setup. From the splitting point, each arm will be fiber-noise compensated. One arm will use an electro-optic modulator to bridge the microwave frequency ($<\SI{20}{\GHz}$) difference between the two clocks. We will monitor this microwave frequency using a microwave reference to detect drifts and oscillating signals potentially emanating from ULDM. The required relative-frequency precision of the microwave reference should be better than $\sim\num{6e-14}$. This precision is achievable with commercially available cold-vapor atomic clocks \cite{marlow2021review}.  

%
\section{Discussion} 

The enhanced sensitivity of the proposed experiment stems from the direct sensitivity of molecular vibration transitions to ULDM that couples to gluons. 
The choice of I$_2^+$ is due to its dense spectrum that allows tuning of the overtone transition to that of an optical atomic clock. Moreover, the homonuclear nature of I$_2^+$ will allow for a long interrogation time due to the decoupling of such molecules from blackbody radiation. 

The dense spectrum of I$_2^+$ imposes an experimental challenge. The clock operation requires the preparation of the molecule in a single quantum state. Due to the homonuclear nature of the molecule, one cannot rely on blackbody-radiation-assisted state preparation \cite{chou2017preparation}. Methods for single-quantum-state preparation for homonuclear molecular ions will rely on quantum-state projection and search algorithms \cite{patterson2018method}.
In addition, the I$_2^+$ molecule has six different total-nuclear-spin-isomer configurations (I=0,1,2,3,4,5). While coherent manipulation of the total-spin-isomer quantum number was suggested~\cite{najafian2020megahertz}, we could also create a total-nuclear-spin-averaged clock.  

To reach the quoted sensitivity, we need to measure the molecular transitions with comparable precision of optical atomic clocks. No fundamental limit exists for achieving such precision in molecules \cite{Carollo:2018arm, hanneke2020optical, kajita2014test}. Moreover, due to their complex spectrum, molecules allow additional tuning mechanisms to enhance precision~\cite{najafian2020megahertz}. However, the systematic-shift analysis for the I$_2^+$ molecule still needs to be performed. 

\emph{Ab-initio} molecular calculations are needed to estimate the magnitude of the coupling of vibrational overtones and the mixing of the electronic excited state ($A\,^2\Pi_{3/2}$) with the electronic ground state. These are crucial for estimating the clock's Rabi frequencies, excited-states lifetimes, and systematic shifts. These molecular calculations are beyond the scope of this work.

The correlation spectroscopy method improves the clock stability by rejecting laser noise. In the case of the Ca$^+$ optical clock, however, we gain only a factor of few in stability compared to the state of the art \cite{huang2022liquid}. This is due to the relatively short excited-state lifetime of Ca$^+$ (\SI{1.16}{\s}). A molecular clock comparison to an optical clock with much longer excited-state lifetimes, such as the $^{27}$Al$^+$ (\SI{20.6}{\s}), $^{87}$Sr (\SI{160}{\s}), or $^{171}$Yb$^+$ (\SI{5}{\year}) optical clocks \cite{Beloy2021,huntemann2016single} will benefit more in terms of stability. Our choice of Ca$^+$ was due to its relatively low transition energy to reduce mixing effects from excited electronic states in the molecule and its upcoming availability in the researchers' institute \cite{akerman2018atomic}. 

A relatively unexplored choice of a clock with a long excited-state lifetime is the Ba$^+$ (\SI{31.2}{\s}) \cite{auchter2014measurement} infrared atomic clock with a transition frequency of $f_{\textrm{Ba}^+}=\SI[separate-uncertainty=false]{170126432449333.3(4)}{\Hz}$~\cite{arnold2020precision}. The iodine's rovibrational transition $f_{35,38\leftarrow 8,39}$ overlaps that of the barium's clock, $f_{\textrm{I}_2^+}-f_{\textrm{Ba}^+}=\SI{-3\pm31}{\GHz}$. The overtone does not overlap the electronic excited \mbox{$A$-state}~[dashed purple arrow in \cref{fig:spectrum}(b)]. The transition's absolute sensitivity, $f_{\mu;35,8}$=\SI{-73.6}{\THz}, is roughly twice smaller than in the case of Ca$^+$ clock comparison. However, the lifetime-limited stability is higher, $\sigma_c(\tau)=\num{8e-16}/\sqrt{\tau/\si{\s}}$. The resulting sensitivity using the Ba$^+$ clock, $\delta\mu/\mu=\num{1.9e-15}/\sqrt{\tau/\si{\s}}$, is $\sim\!2.5$ times better than the sensitivity with the Ca$^+$ clock. 

When performing correlation spectroscopy of two clocks in separated traps, uncorrelated magnetic-field noise can limit the clock's coherence time. Passive and active magnetic-field-noise stabilization can tremendously decrease the magnetic-field-noise amplitude and increase the clock's coherence time \cite{ruster2016long}. The magnetic noise will be correlated when the clocks are located in the same trap. However, for clocks of a different species, the difference in magnetic susceptibility of the two clocks will limit the coherence time. Due to the dense spectrum of molecules, one can find a clock transition with similar susceptibility to that of the atom, and by that, increase the coherence time even more.  

%
\section{Summary}
%

In this letter, we have proposed a search for ULDM-induced oscillations of the nuclear parameters using correlation spectroscopy of I$_2^+$ molecular-ion and Ca$^+$ atomic-ion clocks.
Our scheme exploits the dense rovibrational spectrum of I$_2^+$, which provides transition frequencies that are degenerate with the Ca$^+$ clock transition with a frequency difference in the microwave regime.
Both clocks are interrogated using the same laser, mitigating laser noise.
Our proposed method is adequate for any molecule or molecular ion, diatomic or polyatomic, with a dense spectrum and good clock transitions. 

Our setup's projected reach improves upon the sensitivity of current clock experiments by 1--3 orders of magnitude. 
We have interpreted the projections in terms of ULDM consisting of a scalar field or an axion coupled to gluons. 
We provide sensitivity in the DM mass range $m_\DM \sim \SIrange{e-21}{e-14}{\eV}$. For scalar DM, we exceed the most stringent sensitivity coming from EP tests in the mass range below $m_\phi \lesssim\SI{e-19}{\eV}$.
For axion DM, our sensitivity is competetive to the one from searches for oscillating nEDMs for masses around $m_a \sim \SIrange{2e-17}{2e-15}{\eV}$. In minimal models we cannot probe decay constants that are large enough to avoid a density-induced axion-field profile around Earth.
We have briefly discussed how non-minimal models can modify the Earth-based bound; however, those require radical changes to the \ZN-QCD model, and more work is needed to conclude whether these models are viable.

\bigskip
\begin{acknowledgments}
We would like to thank Abhishek Banerjee, Melina Filzinger, Nitzan Akerman and Roee Ozeri for fruitful discussions and reading of this manuscript, Konstantin Springmann, Stefan Stelzl and Andreas Weiler for critical comments on the amelioration of the Earth bound,
Marco Gorghetto and Surjeet Rajendran for further discussions, as well as Nathaniel Sherrill for providing the data of the Sr/Cs exclusion curves. 
Z.M.\ acknowledges the support of the Diane and Guilford Glazer Foundation Impact Grant for New Scientists, the Center for New Scientists at the Weizmann Institute of Science, the Edith and Nathan Goldenberg Career Development Chair, the Israel Science Foundation (1010/22), and the Minerva Stiftung with funding from the Federal German Ministry for Education and Research.
The work of G.P.\ is supported by grants from the United States-Israel Binational Science Foundation~(BSF) and the United States National Science Foundation~(NSF), the Friedrich Wilhelm Bessel research award of the Alexander von Humboldt Foundation, the German-Israeli Foundation for Scientific Research and Development~(GIF), the Israel Science Foundation~(ISF), the Minerva Stiftung, the SABRA — Yeda-Sela — WRC Program, the Estate of Emile Mimran, and the Maurice and Vivienne Wohl Endowment.
The authors are grateful to the organizers and participants of the program ``Particle \& AMO physicists discussing quantum sensors and new physics 2023'' at the Munich Institute for Astro-, Particle and BioPhysics (MIAPbP) which is funded by the Deutsche Forschungsgemeinschaft (DFG, German Research Foundation) under Germany's Excellence Strategy – EXC-2094 – 390783311.
\end{acknowledgments}

\bibliography{bibliography}

%
\end{document}